\documentclass[aps,reprint,superscriptaddress,nofootinbib,showpacs]{revtex4-1}
\usepackage{amsmath,latexsym,amssymb,hyperref,graphicx}
\usepackage{slashed}
\usepackage{enumerate}
\usepackage{subfigure}
\usepackage{float}

\hypersetup{colorlinks,citecolor= nicegreen,linkcolor= niceblue}
\usepackage{color}
\definecolor{nicered}{rgb}{0.7,0.1,0.1}
\definecolor{nicegreen}{rgb}{0.1,0.5,0.1}
\definecolor{niceblue}{rgb}{0.1,0.1,0.8}

\begin{document}

\title{Fermions localization on de Sitter branes}

\author{Rommel Guerrero}
\author{R. Omar Rodriguez}
\affiliation{Grupo de Investigaciones en F\'isica, Facultad de Ciencias, Escuela Superior Polit\'ecnica de Chimborazo, EC060155-Riobamba, Ecuador}
\affiliation{Departamento de F\'isica, Decanato de Ciencias y Tecnolog\'ia, Universidad Centroccidental Lisandro Alvarado, 3001-Barquisimeto, Venezuela}

\begin{abstract}
 \noindent

Spin 1/2 fields localization on an asymmetric dS${}_4$ scenario, where the brane interpolates between two spacetimes dS${}_5$ and AdS${}_5$, is determined. The bulk spinor is coupled to scalar field of the brane by a  nonminimal Yukawa term compatible with the scenario's geometry. We show that, independently of wall's
thickness, only one massless chiral mode is localized on the wall. The massive chiral modes follow a Schr\"odinger equation, whose potential has a mass gap determined by Yukawa constant, which is a generic property of this system.
The fermions spectrum is defined bellow the gap,  by bound states of both chiralities with the same mass, and above the gap, by a continuous spectrum with local and global resonant modes  of both chiralities and  different mass.

\end{abstract}

\maketitle

\section{Introduction}
A de-Sitter (dS${}_4$) brane corresponds to a dynamic  hypersurface with positive curvature embedded in a higher dimensional spacetime, for example five dimensions.This configuration is phenomenologically interesting  because it is  similar to Friedmann-Robertson-Walker metric \cite{Cvetic:1993xe, Gass:1999gk}. 

In general, the energy density of a brane divides the spacetime into two sectors with different cosmological constants. In the static case, the curvature of the bulk needs to be AdS${}_5$ to confine gravity in four dimensions \cite{Randall:1999vf,Gabadadze:2006jm,Melfo:2010xu}; while, in the dynamic case, the zero mode of gravitational fluctuations is localized on the brane independently of bulk curvature \cite{Kehagias:2002qk,Ghoroku:2003bs,Araujo:2011fm}. The cosmological constant on the dS${}_4$ brane generates, in  the effective potential of the bulk fluctuations, a massive gap that always favors the capture of massless graviton. 

 The brane can be obtained as a pair vacuum solutions to the Einstein equations rigidly connected on a slice of the bulk \cite{Cvetic:1993xe, Battye:2001pb} or as the thin-wall limit of a domain wall, which is a solution to the coupled Einstein-Klein Gordon system, where the scalar field interpolates between the minima of a  self-interaction potential \cite{Guerrero:2002ki,Melfo:2002wd, CastilloFelisola:2004eg}. While in the first case a fine-tuning in the tension of the brane is required to obtain a stable scenario; in the second one the stability is determined by the topological charge of the domain wall even in the thin-wall limit. We will use the second approach to generate a dS${}_4$ brane from the domain wall reported in \cite{Guerrero:2005aw} which  interpolates asymptotically  between a dS${}_5$ and AdS${}_5$ spacetimes.

Now, if our Universe is to be realized on a brane the Standard Model fields should be confined on it, as is the case with the gravity. In particular, fermions localization requires of a nongravitational mechanism because the brane gravity expelled them toward the extra dimension \cite{Bajc:1999mh}.

The matter field localization on a thick domain wall  is a topic that has been studied in several opportunities  
assuming that fermions interact with scalar field via a Yukawa term, $\lambda\bar{\Psi}\Phi(\phi)\Psi$ \cite{Bajc:1999mh,Ringeval:2001cq,Koley:2004at,Melfo:2006hh}.

In the static scenario, the minimal Yukawa coupling, $\Phi=\phi$, leads to a fermions spectrum determined by a zero mode localized in four dimensions and a tower of  massive states moving freely in the bulk. However, in the thin wall limit  the scalar field vanishes and, in consequence, the Yukawa constant, $\lambda$, diverges \cite{Melfo:2006hh}. 

In the dynamic case, and independently of wall thickness,  the minimal Yukawa coupling is insufficient to find a normalizable solution for the  fermions \cite{Melfo:2006hh}. To trap fermions on a thick dS${}_4$ scenario, nonminimal Yukawa coupling is an option that, to our knowledge, has only been considered numerically in two papers: in \cite{Liu:2009ve}, with $\Phi=\phi^k$, a set of resonant massive fermions on the dynamic $Z_2$ wall of \cite{ Wang:2002pka} is found; and in \cite{Liu:2009dwa}, a Yukawa term given by $\Phi=\sin(\phi/\phi_0)\cos^{-\delta}(\phi/\phi_0)$ is considered to obtain  a discrete spectrum of bound fermions on the asymmetric wall reported in \cite{Guerrero:2005aw}, which are expelled progressively toward extra dimensions as the asymmetry increases. 

Under conditions indicated in \cite{Liu:2009dwa}, it is possible to estimate what happens to the bound fermions when the wall thickness is reduced until to obtained the brane. In this case, the effective potential of the fermions is reduced to a like-delta potential, in such a way that all the  bound states, except zero mode, migrate out of the brane, independently of the wall asymmetry. Hence, in the thin-wall limit, the chosen coupling fails to keep the massive states that were confined when the wall had thickness.

In \cite{Barbosa-Cendejas:2015qaa}, a proposal to construct a nonminimal Yukawa coupling in  compatibility with the scenario's geometry was presented. Remarkably, under this mechanism the fermions localization takes place via the warp factor of scenario, as happens with the gravitational fluctuations, and as a consequence it is possible to keep the matter field coupled to the wall even in the limit of zero thickness.  

In this paper we  apply the mechanisms reported in \cite{Barbosa-Cendejas:2015qaa} for coupling fermions on the dS${}_4$ brane obtained from the asymmetric domain wall of \cite{Guerrero:2005aw}. We analytically  find the spinors spectrum, which is determined by a  doubly degenerated set of bound and quasibound modes  whose number and mass increase as the asymmetry of the brane increases. 

The paper is organized as follows: in Sec. \ref{setup}, the fermions's localization mechanism is implemented. In particular, we show that in the fermionic spectrum exists a mass gap between the zero mode and the continuous of the massive modes, which is defined by the Yukawa coupling constant. The constraint on the Yukawa constant to obtain a normalizable solution for the zero mode is also determined. In Sec. \ref{adSitter}, we consider the asymmetric scenario of \cite{Guerrero:2005aw} in order to find the Yukawa coupling compatible with the model; also we show that in the thin wall limit it does not vanish.
In Sec. \ref{fermionspectrum}, the fermionic spectrum is analyzed following the approach of  \cite{Melfo:2010xu}. Because the coupling is preserved in the zero thickness limit,  we analytically find the full spectrum of fermions on the brane. Finally, a summary of our results is presented in Sec. \ref{summary}.

\section{Setup}\label{setup}

Consider the embedding of a dS brane into a 5-dimensional bulk described by a metric with planar-parallel symmetry 
\begin{equation}\label{metric}
    ds^2=e^{2A(z)}\left(-dt^2 +e^{2\beta t} dx^i dx^i+dz^2\right),
\end{equation}
with $i=1, 2, 3$, where  $\beta$ is a positive parameter that  determines the vacuum energy of the four-dimensional spacetime, $\Lambda_4=2\beta^2$.

The domain wall can be obtained as a solution to the coupled Einstein-scalar field system,
\begin{equation}\label{Rab}
    R_{ab}-\frac{1}{2}R g_{ab}=T_{ab}, 
\end{equation}
\begin{equation}\label{Tab}
     T_{ab}=\nabla_a\phi\nabla_b\phi-g_{ab}\left[\frac{1}{2}\nabla_c\phi\nabla^c\phi+V(\phi)\right],
\end{equation}
\begin{equation}\label{phieq}
     \nabla_a\nabla^a\phi=\frac{d V(\phi)}{d\phi}
\end{equation}
with $a, b=0,\dots,4$, where the scalar field $\phi$, which only  depends on the perpendicular coordinate to the wall $z$, interpolates between the minima of the self-interaction potential $V(\phi)$, i.e.
\begin{equation}
    \lim_{z\rightarrow\pm\infty}\phi(z)=\phi_\pm, \qquad\lim_{\phi\rightarrow\phi_\pm}\frac{dV(\phi)}{d\phi}=0
\end{equation}

Following the usual strategy, we find the scalar field and the potential 
\begin{equation}
    \phi^{\prime}(z)^2=-3\left(A^{\prime\prime}-A^{\prime 2}+\beta^2\right),
\end{equation}
\begin{equation}
    V(\phi(z))=-\frac{3}{2}\left(A^{\prime\prime}+3A^{\prime 2}-3\beta^2\right)e^{-2A(z)},\label{potential2}
\end{equation}
and the energy density and pressure 
\begin{equation}
    \rho(z)=\frac{1}{2}\phi^{\prime}(z)^2e^{-2A(z)}+V(\phi(z)),\label{density}
\end{equation}
\begin{equation}
    \text{P}(z)=-\frac{1}{2}\phi^{\prime}(z)^2e^{-2A(z)}+V(\phi(z)),\label{pressure}
\end{equation}
where prime denotes derivative with respect to $z$. 

For a domain wall solution, $e^A$ is an integrable and asymptotically vanishing function, such that,  $\phi^{\prime}e^{-A}\rightarrow 0$ as $|z|\rightarrow\infty$. Hence, (\ref{density}) and (\ref{pressure}) asymptotically tend to $V(\phi_\pm)=\Lambda_{\pm}$, the cosmological constants at each side of the wall.  

We are interested to include fermions on this scenario, and as usual, a Yukawa coupling, $\lambda  \Phi(\phi)\bar{\Psi}\Psi$, must be considered between the spinors and the scalar field of the wall  \cite{Melfo:2006hh}, with $\lambda$ the coupling constant and  $\Phi(\phi)$ a suitable function of $\phi$. 

The behavior of five-dimensional spinor field in the background (\ref{metric}) is determined by Dirac equation  
\begin{equation}
    \Gamma^a\nabla_a\Psi(x,z)=\lambda\Phi(\phi)\Psi(x,z).
\end{equation}
To obtain the coordinate representation of the motion equation, we factor the spinor field as follows
\begin{equation}
    \Psi(x,z)=\Psi_\text{L}(x)u_\text{L}(z)+\Psi_\text{R}(x)u_\text{R}(z),
\end{equation}
where $\Psi_\text{R}^\text{L}(x)\equiv\pm\gamma^5\Psi_\text{R}^\text{L}(x)$ are the chiral states, left (L) and right (R), which satisfy the massive Dirac equation,
\begin{equation}\label{Dirac4}
    i\gamma^\mu\partial_\mu\Psi_\text{R}^\text{L}(x)=m\Psi_\text{L}^\text{R}(x) .
\end{equation}
Thus, for the chiral modes along the additional dimension we find  
\begin{equation}
    \left(\partial_z+2A^\prime(z)\pm\lambda\Phi(\phi)e^{A(z)}\right)u_\text{R}^\text{L}(z)=\pm m\ u_\text{L}^\text{R}(z) . \label{hatu}
\end{equation}

 If additionally we consider  $u(z)=\hat{u}(z)e^{-2A(z)}$, a Schr\"odinger equation for $\hat{u}$ can be obtained 
\begin{equation}
    \left[-\partial_z^2+V_\text{R}^\text{L}(z)\right]\hat{u}_\text{R}^\text{L}(z)=m^2\hat{u}_\text{R}^\text{L}(z)\label{Sch}
\end{equation}
where 
\begin{equation}
    V_\text{R}^\text{L}(z)=\left[\lambda\Phi(\phi(z))e^{A}\right]^2\pm \left[\lambda\Phi(\phi(z))e^{A}\right]^\prime\label{Vpm}
\end{equation}
is the stationary quantum mechanics potential. The solutions of  (\ref{Sch}) correspond to a spectrum of  eigenfunctions  $\hat{u}_\text{R}^\text{L}(z)$ of Schr\"odinger operator, with eigenvalues $m^2$ . 

Let us consider the massless modes, i.e. the states for $m=0$. In this case we find 
\begin{equation}\label{uLR}
    \hat{u}_\text{R}^\text{L}(z)\sim e^{\mp\lambda\int{ \Phi(\phi(z)) e^{A(z)}}\ dz}.
\end{equation}
where we can observe that the confinement of the zero mode is strongly dependent on the Yukawa coupling. In particular, the minimal coupling case, $\Phi(\phi)=\phi$,  has been widely discussed in several opportunities and for a dS${}_4$ scenario it has been proved insufficient to find normalizable solutions for the zero mode \cite{Melfo:2006hh}, because asymptotically the metric factor and scalar field behave like $e^{A(z)}\rightarrow 0$ and $\phi\rightarrow\phi_\pm$, respectively, which implies that $\hat{u}\sim 1$ as $|z|\rightarrow\infty$, in agreement with (\ref{uLR}). 

In the literature there are few  proposals for $\Phi(\phi)$ \cite{Liu:2009dwa, Liu:2009ve} that have allowed to obtain a chiral spectra of bound states on the wall. Here, we consider the following coupling function
\begin{equation}\label{Phi}
e^{A(\phi)}\Phi(\phi)=\Phi_0\ \text{sgn}(\frac{\phi}{\phi_0})\left(\Lambda_4-\frac{1}{3}e^{2A(\phi)} \text{P}(\phi)\right)^{1/2},
\end{equation}
where $\Phi_0=1/(\sqrt{2}\beta)$ is a proportionality constant. The coupling depends on universal wall properties, such as the pressure and the four-dimensional cosmological constant. Under this coupling, the chiral symmetry is broken for $m=0$ and only one of the  chiral states is normalizable, as shown below.

In the coordinate representation, the coupling function (\ref{Phi}) is reduced to \cite{Barbosa-Cendejas:2015qaa} 
\begin{equation}\label{couplingz}
\Phi(z)=\frac{1}{\beta} (e^{-A(z)})^\prime  
\end{equation}
and, according to (\ref{uLR}), the zero mode is determined by
\begin{equation}
    \hat{u}_\text{R}^\text{L}(z)\sim e^{\pm (\lambda/\beta) A(z)} \label{uLR2}.
\end{equation}
Thus, if one of the chiral modes indicated in (\ref{uLR2}) is a square-integrable function,
\begin{equation}
    \int_{-\infty}^{+\infty} |\hat{u}_\text{L}(z)|^2 dz =   \int_{-\infty}^{+\infty} e^{2(\lambda/\beta) A(z)} dz<\infty,
\end{equation}
necessarily $\lambda>\beta/2$. Therefore, the coupling (\ref{couplingz}), unlike the minimal Yukawa coupling, ensures the localization of one of massless chiral state of fermion  independently  of the dS${}_4$ domain wall solution.

Regarding the massive modes, three aspects should be highlighted. First, as was discussed in \cite{Barbosa-Cendejas:2015qaa}, (\ref{couplingz}) leads to 
\begin{equation}
    V_\text{R}^\text{L}(z)=(\lambda/\beta)^2 A^{\prime 2}\pm(\lambda/\beta) A^{\prime\prime},\label{VLR}
\end{equation}
where $V_\text{L}$ is similar to the potential obtained for the gravitational fluctuations, in such a way that the zero mode of gravitation as fermions can coexist simultaneously on the wall. 

Second,  (\ref{VLR}) has a mass gap that separates the continuous modes of the massless state. For a $V(\phi)$ given by (\ref{potential2}), the effective potential can be written as $
    V_\text{R}^\text{L}(z)=\lambda^2\mp (\lambda/\beta) e^{2A}\left[2\rho\pm(\lambda/\beta\mp 1)\text{P}\right]/6
$. Thus, as $|z|\rightarrow\infty$ the second term of $V_\text{R}^\text{L}(z)$ goes to zero and the effective potential takes the value $\lambda^2$.

Finally, notice that (\ref{Sch},\ref{Vpm}) have the form
\begin{equation}\label{SUSY}
    Q^\dag Q\hat{u}_\text{L}=m^2\hat{u}_\text{L}\ ,\quad Q Q^\dag\hat{u}_\text{R}=m^2\hat{u}_\text{R}
\end{equation}
 where $Q= \partial_z+\lambda\Phi(\phi)e^{A}$ and $Q^\dag= -\partial_z+\lambda\Phi(\phi)e^{A}$, i.e, (\ref{Sch},\ref{Vpm}) can be seen as a SUSY quantum mechanics problem \cite{Sukumar:1986bq, ArkaniHamed:1999dc} and, hence, the eigenvalues are positives and the eigenfunctions come in pairs for each eigenvalue,  
 \begin{equation}
\hat{u}_\text{R}(z)=\frac{1}{m}\left(\partial_z+\lambda\Phi(\phi)e^{A}\right) \hat{u}_\text{L}(z)\label{uRuL}
\end{equation}
except for the massless modes. Indeed, this pairing of the mass eigenmodes is required for the existence of massive
fermions satisfying (\ref{Dirac4}).

\section{From thick to thin dS domain walls}\label{adSitter}

A three-parametric domain wall  with positive curvature  in four dimensions can be obtained from (\ref{Rab},\ref{phieq}) where the metric factor, scalar field and self-interaction potential are determined, respectively, by \cite{Guerrero:2005xx}
\begin{eqnarray}
e^{-A(z)}&=&\cosh^{\delta}(\frac{\beta z}{\delta})+\text{sgn}(z)\frac{\alpha\delta}{\beta(1-2\delta)}\cosh^{1-\delta}(\frac{\beta z}{\delta})\nonumber\\&\times&\text{Re}\left[i {}_2F_1\left(\frac{1}{2}-\delta, \frac{1}{2}, \frac{3}{2}-\delta, \cosh^2(\frac{\beta z}{\delta})\right)\right],\label{metricfactor}
\end{eqnarray}
\begin{equation}
    \phi(z)=\phi_0\arctan\sinh(\frac{\beta z}{\delta}),\quad \phi_0=\sqrt{3\delta(1-\delta)}\label{field}
\end{equation}
and
\begin{eqnarray}
V(\phi)&=&\frac{3\beta^2}{2e^{2A(\phi)}}\left[4+\frac{1-\delta}{\delta}\cos^2(\frac{\phi}{\phi_0})\right]\nonumber\\&-&6\beta^2\cos^{2\delta}(\frac{\phi}{\phi_0})\left[\frac{\alpha}{\beta}+\frac{\cos^{-\delta}(\frac{\phi}{\phi_0})\sin(\frac{\phi}{\phi_0})}{e^{A(\phi)} }
\right]^2\label{potential}
\end{eqnarray}
with $e^{A(\phi)}$ defined as
\begin{eqnarray}
    e^{-A(\phi)}&=&\cos^{-\delta}(\frac{\phi}{\phi_0})+\text{sgn}(\frac{\phi}{\phi_0})\frac{\alpha\delta}{\beta(1-2\delta)}\cos^{\delta-1}(\frac{\phi}{\phi_0})\nonumber\\&\times&\text{Re}\left[i {}_2F_1\left(\frac{1}{2}-\delta, \frac{1}{2}, \frac{3}{2}-\delta, \cos^{-2}(\frac{\phi}{\phi_0})\right)\right].
\end{eqnarray}

To avoid coordinate singularities in the sense of \cite{Guerrero:2005aw,Guerrero:2005xx} the following constrain on asymmetry parameter, $\alpha$, must be considered,
\begin{equation}\label{alpha}
    |\alpha|< \frac{\sqrt{\pi}\beta\ (1-2\delta)}{\delta\ \Gamma(\delta)\Gamma(3/2-\delta)\text{Re}\left[(-1)^{\delta}\right]} .
\end{equation}
In agreement with (\ref{alpha}), the $Z_2$ symmetry is relaxed and the scalar field interpolates between two 5-dimensional spacetimes with different cosmological constants; indeed, the wall is the transition region  between an AdS${}_5$ space with cosmological constant
\begin{equation}
    \Lambda_+=-12\alpha\left(\beta+\frac{\alpha \delta\ \text{Re}\left[(-1)^\delta\right]}{\sqrt{\pi}(1-2\delta)}\Gamma(\delta)\Gamma(3/2-\delta)\right)
\end{equation}
and other dS${}_5$  with cosmological constant
    \begin{equation}
   \Lambda_-=+12\alpha\left(\beta-\frac{\alpha \delta\  \text{Re}\left[(-1)^{\delta}\right]}{\sqrt{\pi}(1-2\delta)}\Gamma(\delta)\Gamma(3/2-\delta)\right) .
    \end{equation}

With regard to Yukawa coupling, from (\ref{Phi}) or (\ref{couplingz}), we find
\begin{eqnarray}
    \Phi(\phi)&=&\cos^{-\delta}(\frac{\phi}{\phi_0})\sin(\frac{\phi}{\phi_0})+\frac{\alpha}{\beta} \cos^{\delta}(\frac{\phi}{\phi_0})\nonumber\\
    &+&\text{sgn}(\frac{\phi}{\phi_0})\frac{\alpha\delta}{2\beta(1-2\delta)}\cos^{\delta}(\frac{\phi}{\phi_0}) \sin(\frac{\phi}{\phi_0})\nonumber\\&\times&\text{Re}\left[i {}_2F_1\left(\frac{1}{2}-\delta, \frac{1}{2}, \frac{3}{2}-\delta, \cos^{-2}(\frac{\phi}{\phi_0})\right)\right], \label{Phia}
\end{eqnarray}
which for $\alpha=0$ (when $Z_2$ symmetry is recovered and the bulk curvature is null) is reduced to
\begin{equation}\label{PhiZ2}
    \Phi(\phi)=\cos^{-\delta}(\frac{\phi}{\phi_0})\sin(\frac{\phi}{\phi_0}) .
\end{equation}
So, (\ref{Phia}) and (\ref{PhiZ2}) are compatible, respectively, with the absence and presence of the $Z_2$ symmetry of spacetime.

In \cite{Liu:2009dwa} a Yukawa term supported on (\ref{PhiZ2}) is used to couple fermions on the dS${}_4$ wall (\ref{metricfactor}, \ref{field}, \ref{potential}) for any value of $\alpha$ and as a result the number of fermions captured by the coupling decreases with the increase of $\alpha$. In our opinion, this effect is generated by the incompatibility between the Yukawa coupling (\ref{PhiZ2}) and the asymmetry of the spacetime.

\begin{figure}[H]
\begin{center}
 \includegraphics[width=8.2cm,angle=0]{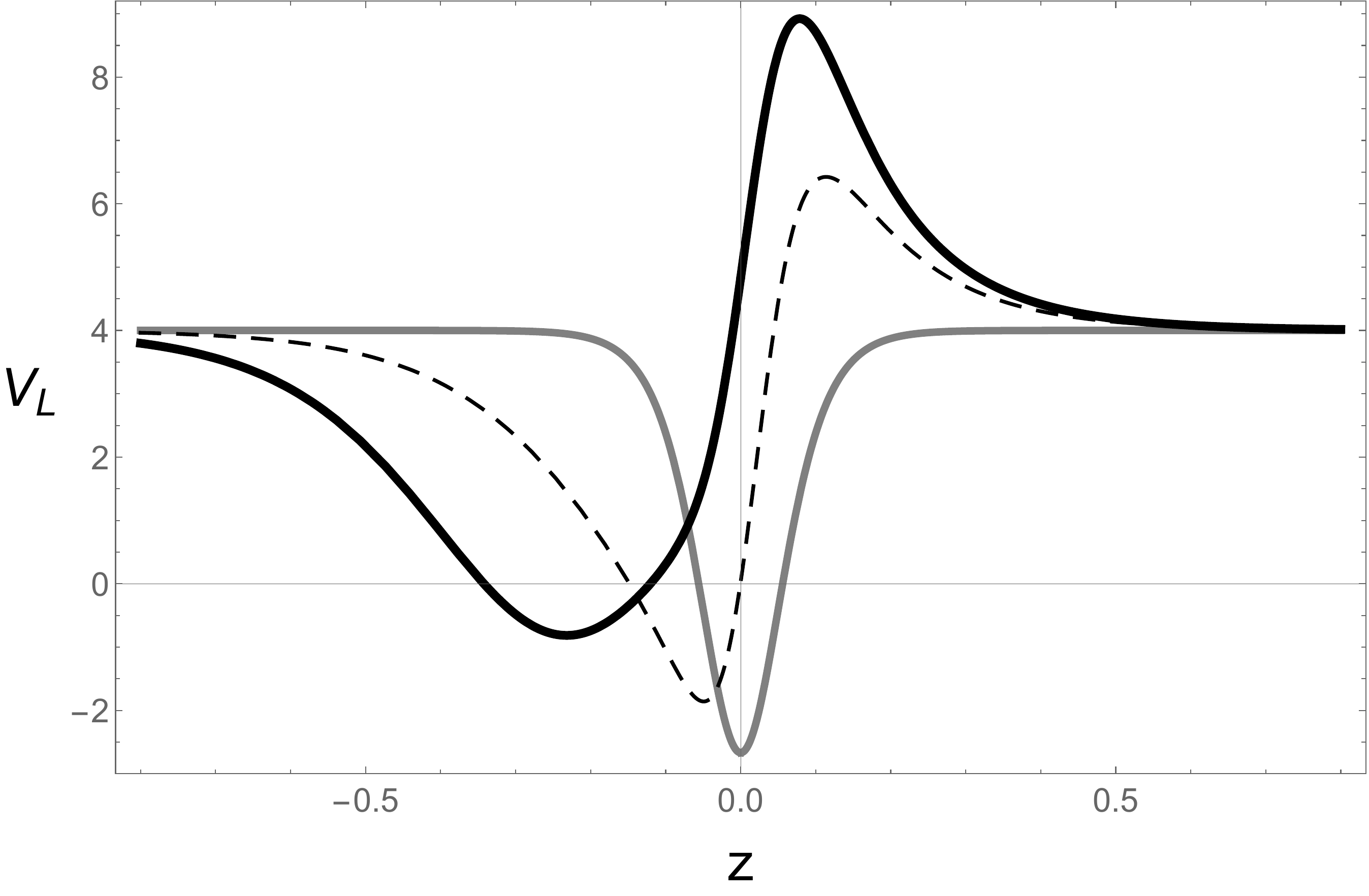}
 \includegraphics[width=8.2cm,angle=0]{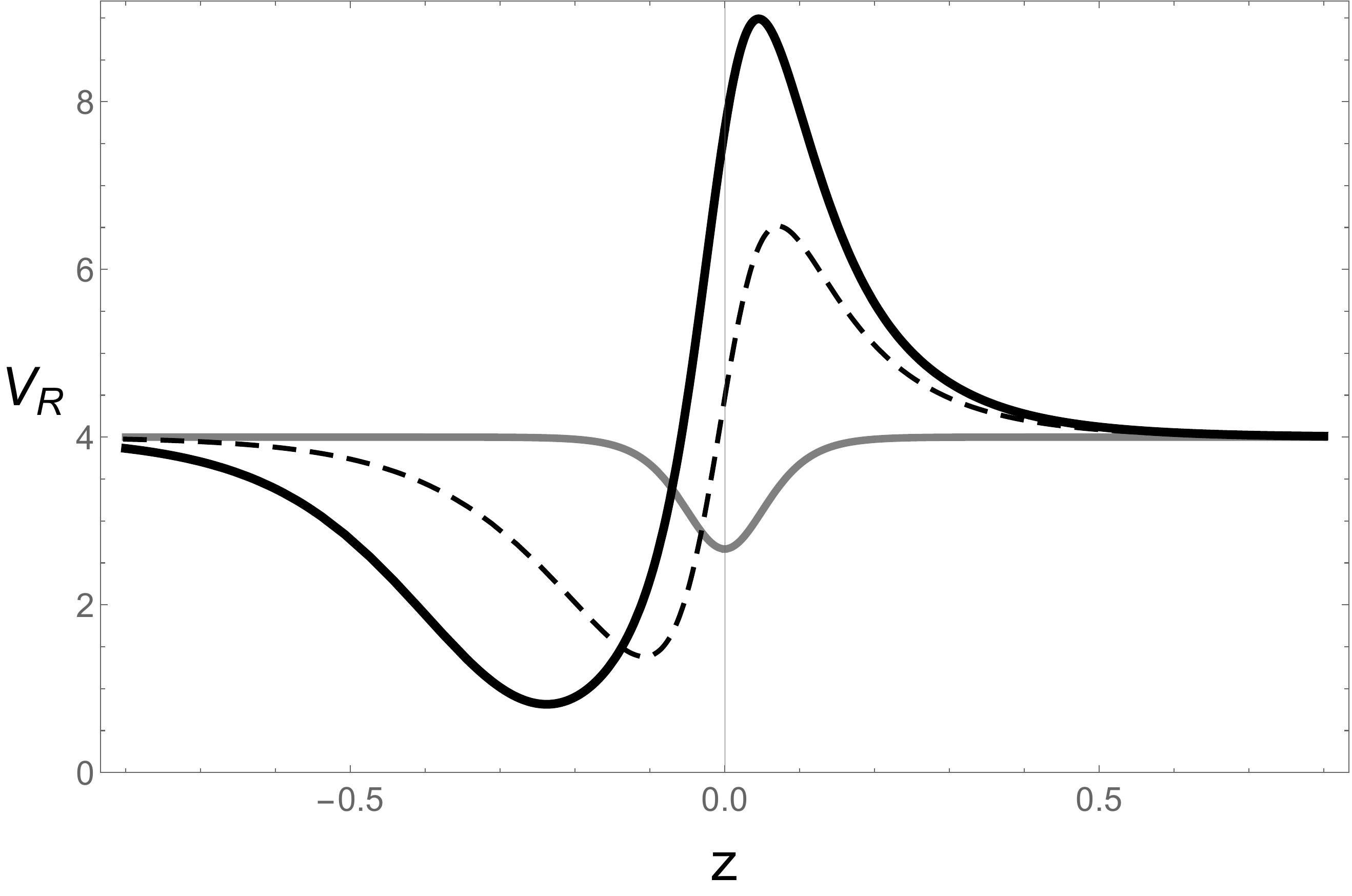}
 \caption{Plots of potentials $V_\text{L}$ (top panel) and $V_\text{R}$ (bottom panel) in the dS thick wall, for $\alpha=0$ (gray line), $\alpha=3$ (dashed line) and $\alpha=6$ (black line).}
\label{VLVRthick}
\end{center}
\end{figure}

According to the Yukawa coupling defined by (\ref{Phia}) and from (\ref{Vpm}) an asymmetric volcano potential is obtained. In Fig.\ref{VLVRthick}, the effective potentials $V_{\text{L}}$ (top panel) and $V_\text{R}$ (bottom panel) are shown for different $\alpha$. In both cases  the eigenfunctions are determined, below the gap ($m<\lambda$), by a set of chiral partner of bound states and, above the gap ($m>\lambda$), by a set of the chiral partners of continuous modes, which  propagate freely for the bulk. Additionally, due to the absence of $Z_2$ symmetry in the potentials, resonance
modes in the spectrum of fluctuations are expected \cite{Gabadadze:2006jm,Melfo:2010xu,Araujo:2011fm}. 

Notice that Fig.\ref{VLVRthick} also shows how the area of the well increases with the increase of the asymmetry; thus, it is expected that the number of the bound states to the well also increases with the asymmetry, unlike the effect reported in \cite{Liu:2009dwa}.

For a domain wall to be a realization of our Universe the thickness of the wall must be infinitely thin.  So, we will analytically determine  the full spectrum of the chiral fermions in the thin wall limit of scenario (\ref{metricfactor}, \ref{field}, \ref{potential}). 

By taking $\delta\rightarrow 0$, the warp factor (\ref{metricfactor}) is reduced to
\begin{equation}\label{MetricFactordelta0}
    e^{-A(z)}=e^{\beta |z|}+\frac{\alpha}{\beta}\sinh(\beta z)\ ,\qquad |\alpha|<2\beta  
\end{equation}
and the spacetime behaves asymptotically as two subspaces with different cosmological constants at either side of the brane
\begin{equation}
    \Lambda_-=6\alpha(2\beta-\alpha)\ ,\qquad \Lambda_+=-6\alpha(2\beta+\alpha)
\end{equation}
in correspondence with the energy density
\begin{equation}
    \rho(z)=\Lambda_-\Theta(-z)+6\beta\delta(z)+\Lambda_+\Theta(z).
\end{equation}
We leave in the Appendix the technical details associated to (\ref{MetricFactordelta0}).

Now, it should be noticed that as $\delta\rightarrow 0$ the
scalar field $\phi$ vanishes everywhere  while
\begin{equation}
    \Phi(z)=\text{sgn}(z) e^{\beta |z|}+\frac{\alpha}{\beta}\cosh(\beta z) .
\end{equation}
Hence, confined fermions on the brane are expected.

\section{Fermions spectrum}\label{fermionspectrum}

\begin{figure}[H]
\begin{center}
\includegraphics[width=8.5cm,angle=0]{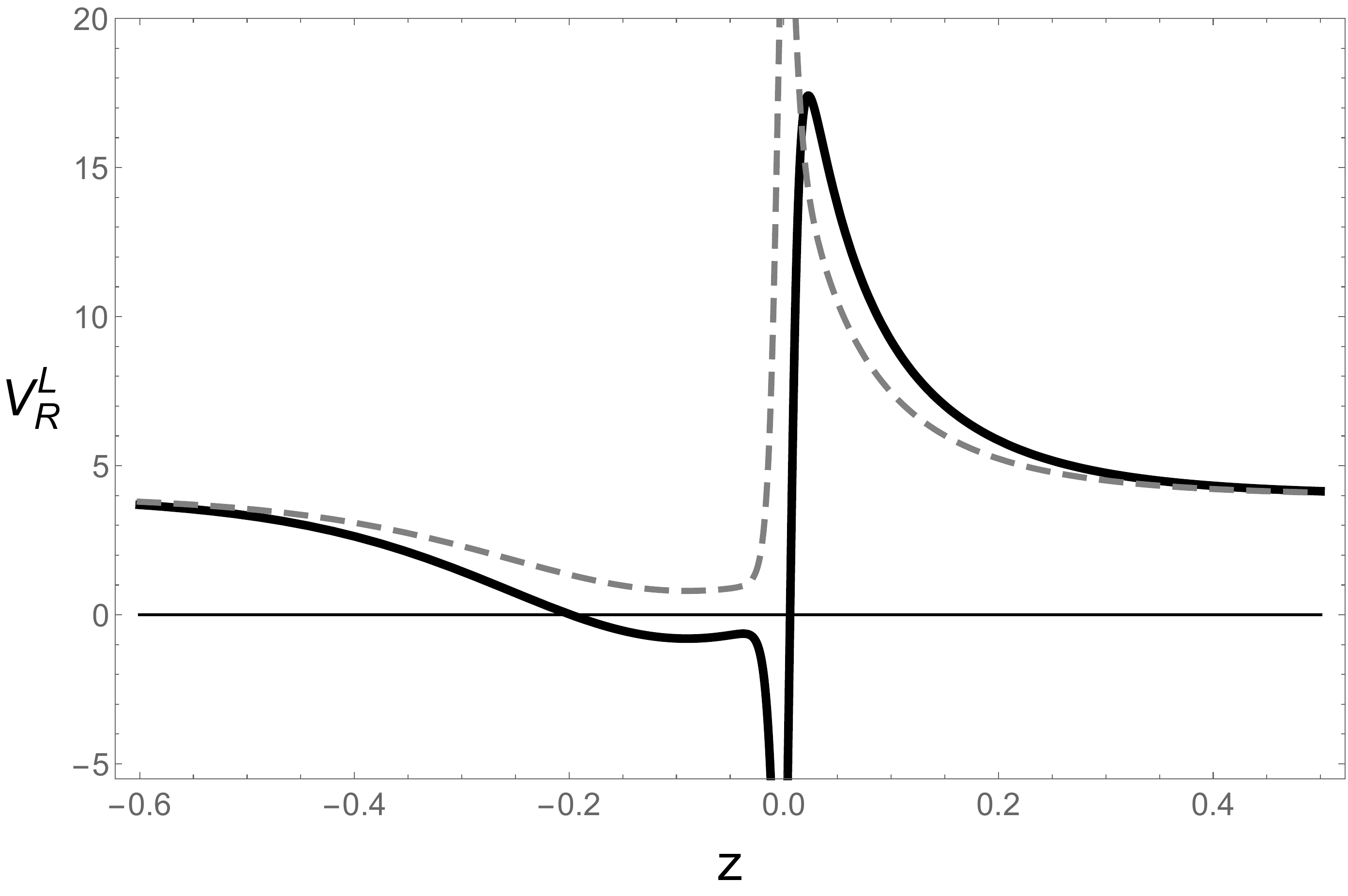}
 \caption{Plots of potentials $V_\text{L}$ (solid line) and $V_\text{R}$ (dashed line) in the dS${}_4$ thin wall.}
\label{VLRthickness}
\end{center}
\end{figure}

In the zero thickness limit, (\ref{VLR}) takes the form 
\begin{eqnarray}
V_\text{R}^\text{L}(z)&=&\mp2\lambda \delta(z)+ \lambda^2\nonumber\\\nonumber\\
&&-4(\lambda/\beta)(\lambda/\beta\pm1) \left(\frac{C_- \beta e^{\beta z}}{1+C_-^2 e^{2\beta z}}\right)^2 \Theta(-z)\nonumber\\\nonumber\\
&&-4(\lambda/\beta)(\lambda/\beta\pm1)\left(\frac{C_+ \beta e^{-\beta z}}{1+C_+^2e^{-2\beta z}}\right)^2 \Theta(z)\nonumber\\\label{VQMdelta}
\end{eqnarray}
with
\begin{equation}
C_{\pm}=\sqrt{\frac{\beta^2}{\Lambda_\pm/6}}-\sqrt{\frac{\beta^2}{\Lambda_\pm/6}-1};
\end{equation}
where the $\delta(z)$ function is associated with either an infinite well for $V_\text{L}$ or an infinite barrier for $V_\text{R}$. In addition, for $z<0$ the  potentials exhibit a smooth well where it is possible to find bound massive states for both chirality modes. The Fig.\ref{VLRthickness} shows a regularized version of $V_\text{R}^\text{L}$ as $\delta\rightarrow 0$.

The eigenvalue problems (\ref{Sch},\ref{VQMdelta}) are similar to those considered in \cite{Melfo:2010xu, Araujo:2011fm} where two wave functions are identified for each $m^2$, such that, while one is transparent to the brane, $\hat{u}^c_{m}$, the other one, $\hat{u}^d_{m}$, is scattered for it. We have
\begin{equation}\label{match-uc}
    \hat{u}^c_{m-}(0)=\hat{u}^c_{m+}(0)=0 ,
\end{equation}
\begin{equation}\label{match-duc}
    \frac{d}{dz}\hat{u}^c_{m-}(0)-\frac{d}{dz}\hat{u}^c_{m+}(0)=0
\end{equation}
for the transparent states  to the brane; and the following conditions for the modes sensible to the brane 
\begin{equation}\label{match-ud}
    \hat{u}^d_{m-}(0)=\hat{u}^d_{m+}(0) ,
\end{equation}
\begin{eqnarray}
   && \frac{d}{dz}\hat{u}^{\text{L}d}_{m-}(0)-\frac{d}{dz}\hat{u}^{\text{L}d}_{m+}(0)=2 \lambda\ \hat{u}^{\text{L}d}_{m}(0),\label{match-dudL}\\
  &&  \frac{d}{dz}\hat{u}^{\text{R}d}_{m-}(0)-\frac{d}{dz}\hat{u}^{\text{R}d}_{m+}(0)=-2 \lambda\  \hat{u}^{\text{R}d}_{m}(0)\label{match-dudR}.
\end{eqnarray}

Rnonegarding orthogonality relationship, for $m\leq \lambda$ the bound states satisfy
\begin{equation}\label{orto-u}
    \int{\hat{u}^{*i}_{m^\prime}(z)\hat{u}^j_{m}(z)dz}=\delta^{ij}\delta_{mm^\prime} ,\quad i, j=c, d;
\end{equation}
and for $m>\lambda$ a similar relationship is obtained changing $\delta_{m m^\prime}$ by $\delta(m-m^\prime)$ for continuous modes. In this last case a procedure to regularize the eigenfunctions must be done, see \cite{Callin:2004py} for details.  

\subsection{Bound states: $m\leq\lambda$}

Below the gap, $0< m\leq \lambda$, the eigenfunctions of (\ref{Sch}, \ref{VQMdelta}) are determined by a discrete tower of massive modes constrained for the boundary conditions $\hat{u}_{\text{L}m}(\pm\infty)=
\hat{u}_{\text{R}m}(\pm\infty)=0$. Defining 
\begin{equation}
\mu\equiv\sqrt{\lambda^2-m^2},
\end{equation}
and 
\begin{equation}
F_{\pm}^{\text{L}}\equiv {}_2F_1\left[\frac{\lambda}{\beta}+1, -\frac{\lambda}{\beta}, 1+ \frac{\mu}{\beta}; \left(1+ \frac{1}{C_\pm^2}e^{\pm 2\beta z}\right)^{-1}\right] ,
\end{equation}
\begin{equation}
F_{\pm}^{\text{R}}\equiv {}_2F_1\left[-\frac{\lambda}{\beta}+1, 
 \frac{\lambda}{\beta}, 1+ \frac{\mu}{\beta}; \left(1+ \frac{1}{C_\pm^2}e^{\pm 2\beta z}\right)^{-1}\right] ,
\end{equation}
in concordance with asymptotic conditions, we have
\begin{eqnarray}
\hat{u}_{m}(z)=N_m \left[A e^{\mu z}F_{-}\Theta(-z)+e^{-\mu z}F_{+}\Theta(z)\right],
\label{modosmasivos1}
\end{eqnarray}
where $N_m$ and $A$ are the normalization and integration constants  to be determined from the integrability conditions. Notice that, for  $\hat{u}^c_{m}$ the trivial solution is obtained while for $\hat{u}^d_{m}$ the corresponding constants can be found from (\ref{match-ud}) and (\ref{orto-u}). Furthermore, (\ref{match-dudL},\ref{match-dudR}) lead to a restriction on the values of  $m/\lambda$ determined by the transcendental equation $y_1=y_2$, where
\begin{eqnarray}\label{teqL}
y_{1\text{L}}&=& 2\left(\lambda+\mu\right)\nonumber\\
&&+\frac{\lambda\left(\alpha-2\beta\right){}_2F_1\left[1-\frac{\lambda}{\beta},1+\frac{\lambda}{\beta},1+\frac{\mu}{\beta};\frac{\alpha}{2\beta}\right]}{ {}_2F_1\left[-\frac{\lambda}{\beta},1+\frac{\lambda}{\beta},1+\frac{\mu}{\beta};\frac{\alpha}{2\beta}\right]\beta} , \\
y_{2\text{L}}&=&\frac{\lambda\left(\alpha+2\beta\right){}_2F_1\left[1-\frac{\lambda}{\beta},1+\frac{\lambda}{\beta},1+\frac{\mu}{\beta};-\frac{\alpha}{2\beta}\right]}{{}_2F_1\left[-\frac{\lambda}{\beta},1+\frac{\lambda}{\beta},1+\frac{\mu}{\beta};-\frac{\alpha}{2\beta}\right]\beta} ,
\end{eqnarray}
for the left modes, and 
\begin{eqnarray}\label{teqR}
y_{1\text{R}}&=& 2\left(-2\beta+3\lambda+\mu\right)\nonumber\\
&&+\frac{(\lambda-\beta)\left(\alpha-2\beta\right){}_2F_1\left[2-\frac{\lambda}{\beta},\frac{\lambda}{\beta},1+\frac{\mu}{\beta};\frac{\alpha}{2\beta}\right]}{{}_2F_1\left[1-\frac{\lambda}{\beta},\frac{\lambda}{\beta},1+\frac{\mu}{\beta};\frac{\alpha}{2\beta}\right]\beta} ,\nonumber\\ \\
y_{2\text{R}}&=&\frac{(\lambda-\beta)\left(\alpha+2\beta\right){}_2F_1\left[2-\frac{\lambda}{\beta},\frac{\lambda}{\beta},1+\frac{\mu}{\beta};-\frac{\alpha}{2\beta}\right]}{{}_2F_1\left[1-\frac{\lambda}{\beta},\frac{\lambda}{\beta},1+\frac{\mu}{\beta};-\frac{\alpha}{2\beta}\right]\beta} ,\nonumber\\
\end{eqnarray}
for the right modes.

\begin{figure}[H]
\begin{center}
\includegraphics[width=8.2cm,angle=0]{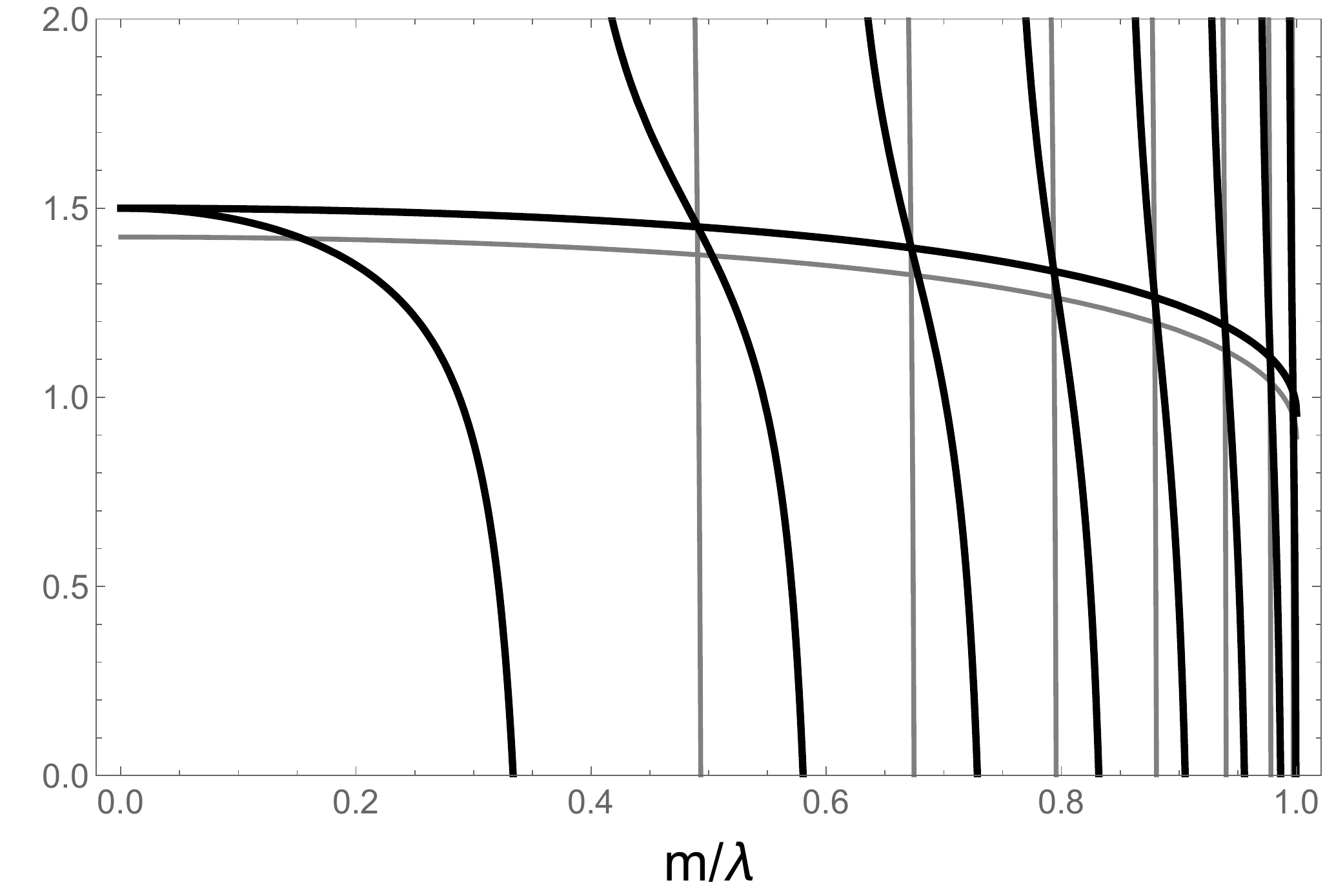}
 \caption{A graphical solution of equations $y_{1\text{L}}=y_{2\text{L}}$ (black) and $y_{1\text{R}}=y_{2\text{R}}$ (gray) for the mass of seven bound states of chiral fermion in a dS spacetime with parameters $\beta=\alpha=5$ and  Yukawa constant $\lambda=15\beta$. 
 }\label{boundedmass}
\end{center}
\end{figure}

In Fig. \ref{boundedmass} a numerical solution for the transcendental equations is shown, it is observed that both equations have the same solution, the interception of $y_1$ and $y_2$, in each mode, occurs for the same values of mass. Thus, the bound modes supported for $V_\text{L}$ and $V_\text{R}$ share  the same eigenvalues. In agreement with (\ref{SUSY}), $V_\text{L}$ and $V_\text{R}$ are supersymmetric partner potentials \cite{Sukumar:1986bq, ArkaniHamed:1999dc}; therefore,  the result showed in the graphic for the eigenvalues is to be expected.

This tower of discrete mass is determined  between the minimum of $V_\text{R}$ and the gap associated to $V^\text{L}_\text{R}$, i.e.,  $\lambda^2>m^2>\lambda\beta$; in fact,  applying the harmonic oscillator approximation around the minimum of  $V_\text{L}$, we found 
\begin{equation}
    m_n^2\simeq (2n+1)\beta\sqrt{\lambda(\lambda-\beta)}+\lambda\beta , \label{massn}
\end{equation}
with $n=0, 1, 2, \dots$ such that $n<(\sqrt{\lambda(\lambda-\beta)}-\beta)/2\beta$.
On the other hand, the number of bound modes in good approximation can be estimated by
\begin{equation}
    {\cal N}\simeq 1+\mathbb{E}\left[\frac{2}{5}\left(1+\sqrt{z_0^2 V_0}\right)\left(1-e^{-\alpha/(2\beta)}\right)\right] \label{N1}
\end{equation}
where $\mathbb{E}[\xi]$ is the integer part of $\xi$, and $z_0$ and $V_0$ are given by
\begin{equation}
    z_0=\frac{1}{4\beta}\ln\frac{19non\lambda+20\beta+2\sqrt{10}\sqrt{(\lambda+\beta)(9\lambda+10\beta)}}{19\lambda+20\beta-2\sqrt{10}\sqrt{(\lambda+\beta)(9\lambda+10\beta)}} \label{z0}
\end{equation}
and 
\begin{equation}
  V_0 = 2 \lambda^2+\lambda \alpha( \lambda+\beta)(2\beta-\alpha)/\beta^2 .\label{V0}
\end{equation}

Notice that for $\alpha=0$, when the scenario exhibits $Z_2$ symmetry, the potential is like delta and consistently ${\cal N}=1$. When the asymmetry increases, $\alpha\rightarrow 2\beta$, the number of bound states increases until it reaches the  saturation value which is given by
\begin{equation}
    {\cal N}\simeq 1+\mathbb{E}\left[\frac{2}{5}\left(1+\sqrt{2}\ z_0 \lambda\right)\left(1-e^{-1}\right)\right] .
\end{equation}
Furthermore, from (\ref{N1}-\ref{V0}) we can see,  consistently, that  the amount of bound states increases as the Yukawa coupling increases.

\subsection{Resonant states: $m>\lambda$}
Above the gap, $m>\lambda$, we find a continuous tower of massive modes confronting to the potential (\ref{VQMdelta}). Considering 
\begin{equation}
\mu\equiv\sqrt{m^2-\lambda^2}
\end{equation}
and
\begin{equation}
F_{\pm}^{\text{L}}\equiv {}_2F_1\left[\frac{\lambda}{\beta}+1, -\frac{\lambda}{\beta}; 1- i\frac{\mu}{\beta}; \left(1+ \frac{1}{C_\pm^2}e^{\pm 2\beta z}\right)^{-1}\right] ,
\end{equation}
\begin{equation}
F_{\pm}^{\text{R}}\equiv {}_2F_1\left[-\frac{\lambda}{\beta}+1, \frac{\lambda}{\beta}; 1- i\frac{\mu}{\beta}; \left(1+ \frac{1}{C_\pm^2}e^{\pm 2\beta z}\right)^{-1}\right] ,
\end{equation}
the modes can be written, for $z<0$, as
\begin{eqnarray}
\hat{u}_{m-}(z)&=&N_m\left[A_- \left(e^{-i \mu z}F_{-}+e^{i \mu z}F^*_{-}\right)\right.\nonumber\\&&\left. -i B_-\left(e^{-i \mu z}F_{-}-e^{ i \mu z}F^*_{-}\right)\right],\label{modosmasivos2menos}
\end{eqnarray}
and for $z>0$, as
\begin{eqnarray}
\hat{u}_{m+}(z)&=&N_m\left[A_+ \left(e^{i \mu z}F_{+}+e^{- i \mu z}F^*_{+}\right)\right.\nonumber\\&&\left.- i\left(e^{i \mu z}F_{+}-e^{- i \mu z}F^*_{+}\right)\right],\label{modosmasivos2mas}
\end{eqnarray}
where $N_m$ and $A_\pm, B_-$ are constants which can be fixed using the integrability  conditions listed previously. 

In this case, nontrivial solutions for (\ref{Sch},\ref{VQMdelta}), $\hat{u}^c_{m}$ and $\hat{u}^d_{m}$, are obtained, in such a way that for each chiral state there are two eigenfuntions sharing the same eigenvalue; that is, the Kaluza-Klein tower is determined by chiral states, each  of them doubly degenerated.

\begin{figure}[H]
\begin{center}
\includegraphics[width=8.8cm,angle=0]{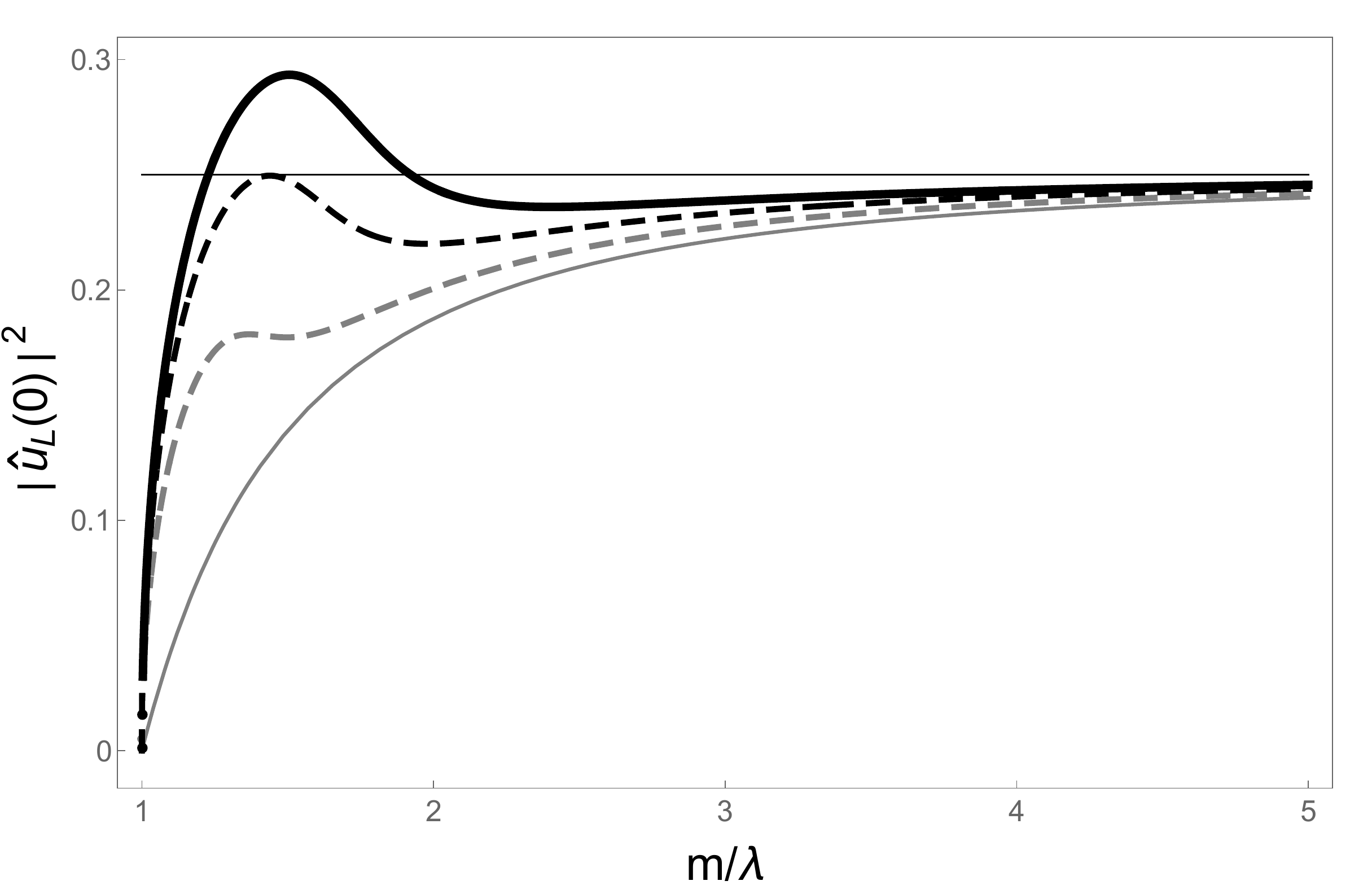}
\caption{Plots of left resonant modes for different asymmetric scenarios for fixed Yukawa constant and $|\Lambda_-/\Lambda_+|=1$ (gray line), $|\Lambda_-/\Lambda_+|=1/2$ (gray dashed line), $|\Lambda_-/\Lambda_+|=2/5$ (black dashed line), $3/10$ (black line).}\label{uL}
\includegraphics[width=8.8cm,angle=0]{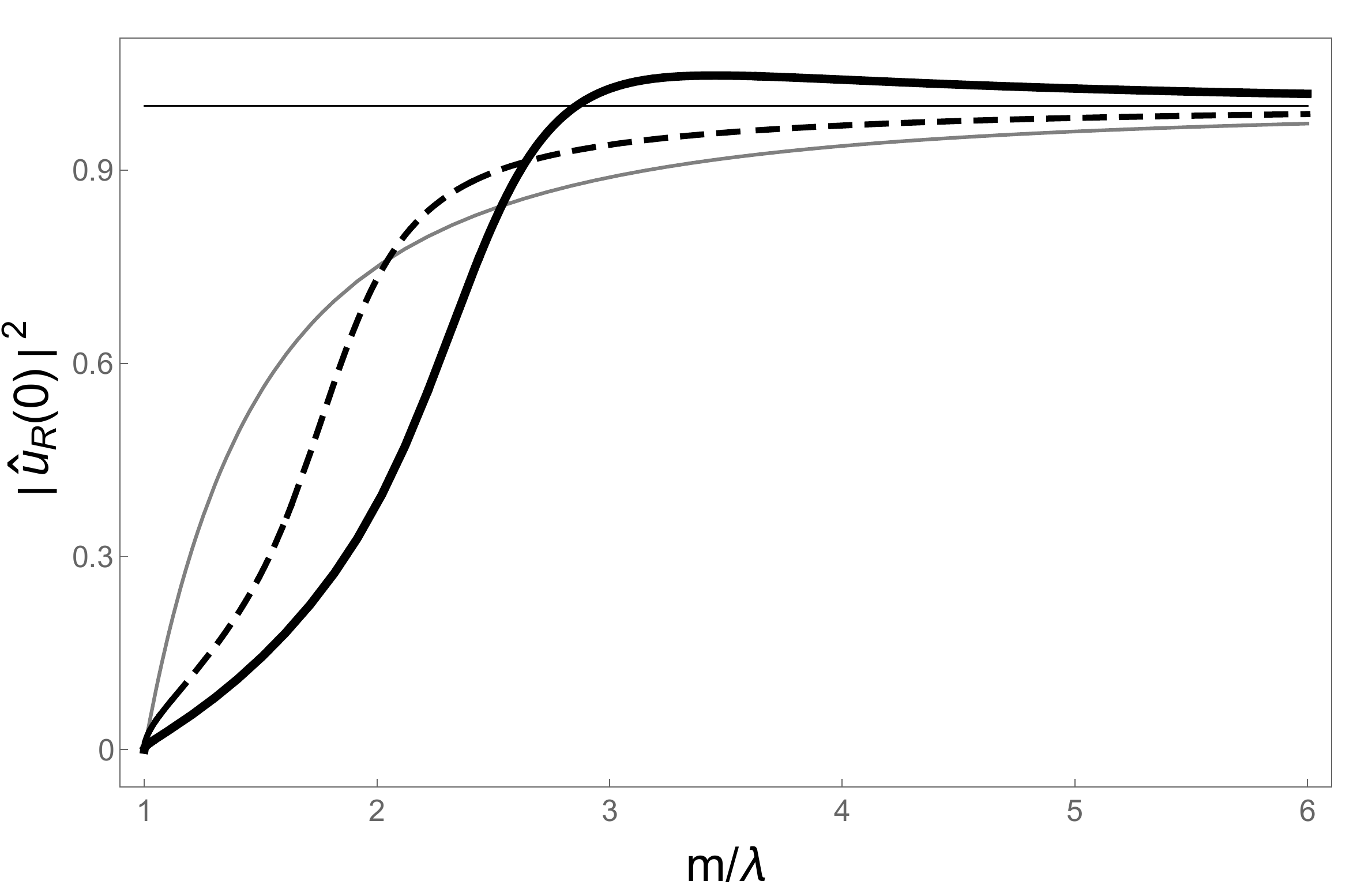}
\caption{Plots of right resonant modes for different asymmetric  scenarios for fixed Yukawa constant and  $|\Lambda_-/\Lambda_+|=1$ (gray line), $|\Lambda_-/\Lambda_+|=1/5$ (dashed line) and $|\Lambda_-/\Lambda_+|\sim 0$ (black line).}\label{uR}
\end{center}
\end{figure}

The existence of resonant massive modes should be highlighted at $z=0$. The behavior of $|\hat{u}_m^d(0)|^2$  for different values of
$|\Lambda_-/\Lambda_+|\leq 1$ is shown in Figs. \ref{uL} and \ref{uR} for the left and right modes respectively. In any case, notice that the resonant mass increases as the asymmetry increases.

Fig.\ref{uL}, in contrast to Fig.\ref{uR}, shows local resonant states for $\hat{u}_L$  which appear when the asymmetry is in correspondence with  $3/5>|\Lambda_-/\Lambda_+|\geq 2/5$.  In particular,   $|\Lambda_-/\Lambda_+|\sim 2/5$ is a critical case where the  state resonant is a light state with  the same probability as heavy ones. Global resonant modes are meaningful for $|\Lambda_-/\Lambda_+|<2/5$. 

The mass of the resonant states can be roughly estimated. For the left mode we have
\begin{equation}
    m_{\text{res}}^{\text{L}}\sim \left[\left(\lambda^2+ \frac{|\Lambda_+|}{6\beta^2}(\lambda+\beta )^2\right)\left(\lambda^2-\frac{\Lambda_-}{6}\right)\right]^{1/4} \label{massrL}
\end{equation}
while  for the right mode, as long as $|\Lambda_-/\Lambda_+|\sim 0$, it is  given by 
\begin{equation}
    m_{\text{res}}^{\text{R}}\sim 3 \lambda\label{massrr}.
\end{equation}
Comparing both results 
\begin{equation}
    m_{\text{res}}^{\text{L}}\sim \frac{1}{3}\left[9+\frac{8\beta}{\lambda}\left(2+\frac{\beta}{\lambda}\right)\right]^{1/4}m_{\text{res}}^{\text{R}}
\end{equation}
 it is deduced that the left and right modes  resonate for different mass. Therefore, for the resonant modes the chiral symmetry of the fermions is quasibroken.

\section{Summary} \label{summary}

We determined  the fermions spectrum on a dS${}_4$ brane embedded in a spacetime without reflection symmetry, from matter fields coupled to the dynamic domain wall of \cite{Guerrero:2005aw}: we coupled the spinors to the scalar field of the wall by the non-minimal Yukawa term (\ref{Phi}), the tower of  fermions was obtained in the thin-wall limit where the Yukawa interaction considered does not vanish and the domain wall becomes a dS${}_4$ brane.

We showed that in the dS${}_4$ brane the nonminimal Yukawa coupling leads an effective potential that supports a spectrum of fermions determined by: a massless broken chiral fermion localized in four dimensions, a set of massive  bound states and a continuum tower of massive bulk states.   

The zero mode of the spectrum is separated from the continuous one by a mass gap which it is defined by the coupling constant in such a way that the number of massive modes increases with intensity of coupling. On the other hand, the number of  the bound states also increases when the asymmetry parameter of scenario increments as a consequence of compatibility between the nonconventional Yukawa coupling and the spacetime geometry.

With regard to the continuous modes, each chiral state is doubly degenerate: while one of them is transparent to the brane the other one is scattered by it. Among scattered ones, chiral resonant modes with different mass can be found, i.e., one of chiral states of resonant fermion is more likely to be on the brane. Therefore, for a four-dimensional observer the chiral symmetry of the resonant spinor is broken.

\section{Acknowledgements}
This work was supported by IDI-ESPOCH under the project titled {\it Domain walls and their gravitational effects.} The authors wish to thank Francisco Carreras for his collaboration to complete this paper. 

\section{Appendix}
Consider the spacetime ($\mathbb{R}^5$
, g), with the metric
\begin{equation}\label{Ap1}
{g}_{ab}=e^{2A(z)}\left(-dt_a dt_b+e^{2\beta t} dx^i_a dx^i_b+dz_adz_b\right)
 \end{equation}
where the metric factor 
\begin{equation}\label{Ap2}
  e^{-A(z)}=  e^{-A_-(z)}\Theta(-z)+e^{-A_+(z)}\Theta(z)
\end{equation}
and
\begin{eqnarray}\label{FM+}
e^{-A_\pm(z)}&=&\cosh^{\delta}(\frac{\beta z}{\delta})\pm\frac{\alpha\delta}{\beta(1-2\delta)}\cosh^{1-\delta}(\frac{\beta z}{\delta})\nonumber\\&\times&\text{Re}\left[i {}_2F_1\left(\frac{1}{2}-\delta, \frac{1}{2}, \frac{3}{2}-\delta, \cosh^2(\frac{\beta z}{\delta})\right)\right]\quad
\end{eqnarray}
being ${}_2F_1$ the Gauss hypergeometric function defined under the following restrictions
\begin{itemize}
\item If $c\notin \mathbb{Z}^-$ and $|\xi|<1$, the function ${}_2F_1$ is given by 
\begin{equation}\label{2F1S}
{}_2F_1\left(a, b; c; \xi\right)=\sum_{n=0}^\infty\frac{(a)_n (b)_n}{(c)_n}\ \frac{\xi^n}{n!}\ ,
\end{equation}
with $(a)_n=\Gamma(a+n)/\Gamma(a)$ the Pochhammer's symbol. 
\item If $a-b\notin \mathbb{Z}$ and $|\xi|>1$, the function ${}_2F_1$ is given by
\begin{eqnarray}\label{2F1>1}
&&{}_2F_1\left(a, b; c; \xi\right)=\frac{\Gamma(c)\Gamma(b-a)}{\Gamma(c-a)\Gamma(b)}\left(-\xi\right)^{-a}\nonumber\\&&\qquad\qquad\qquad\times{}_2F_1\left(a,c-b;a-b+1;\frac{1}{\xi}\right)\nonumber\\
&&\qquad\qquad\qquad+\frac{\Gamma(c)\Gamma(a-b)}{\Gamma(a)\Gamma(c-b)}\left(-\xi\right)^{-b}\nonumber \\&&\qquad\qquad\qquad\times{}_2F_1\left(b,c-a;b-a+1;\frac{1}{\xi}\right).
\end{eqnarray}
\end{itemize}
The solution (\ref{Ap1}, \ref{Ap2}, \ref{FM+}) represents a five-dimensional dynamic domain wall  with $\delta$ playing the role of the wall's thickness. Next, we will examine the thin wall limit $\delta\rightarrow 0$, of this solution.

We first consider  in (\ref{FM+}) the linear transformation
\begin{equation}\label{2F1LT}
  {}_2F_1\left(a, b; c; \xi\right)=(1-\xi)^{-a}{}_2F_1\left(a, c-b; c;  \frac{\xi}{\xi-1}\right) ,
\end{equation}
where $\xi=\cosh^2(\beta z/\delta)$, $a=1/2-\delta$, $b=1/2$ and $c=3/2-\delta$. On the other hand, in this case ${}_2F_1(a, c-b; c;  \xi/(\xi-1))$ is given by (\ref{2F1>1}) and, therefore, Eq. (\ref{2F1LT}) can be written as
\begin{eqnarray}\label{Identity}
&&{}_2F_1\left(a, b; c; \xi\right)=\frac{\Gamma(c)\Gamma(b-a)}{\Gamma(c-a)\Gamma(b)}\left(1-\xi\right)^{-a}\nonumber\\&&\qquad\qquad\qquad\times{}_2F_1\left(a,c-b;a-b+1;\frac{1}{1-\xi}\right)\nonumber\\
&&\qquad\qquad\qquad+\frac{\Gamma(c)\Gamma(a-b)}{\Gamma(a)\Gamma(c-b)}\left(1-\xi\right)^{-b}\nonumber \\&&\qquad\qquad\qquad\times{}_2F_1\left(b,c-a;b-a+1;\frac{1}{1-\xi}\right) .
\end{eqnarray}
In this representation is easy to check the convergence of the metric factor for a small thickness of the wall, because  $|(1-\xi)^{-1}|=\text{csch}^2(\beta z/\delta)\sim 0$ as $\delta\sim 0$, and in consequence, the hypergeometric functions in (\ref{Identity}) are defined by the power series (\ref{2F1S}). Thus, for $\delta\sim 0$ the function $e^{-A}$ is given by
\begin{eqnarray}
e^{-A_\pm(z)}&\simeq&\frac{\alpha}{2\beta}\left[2^{\mp\delta}e^{\beta z}\mp(-2)^{\pm\delta}e^{-\beta z}\right]\nonumber\\&&+2^{-\delta}e^{\pm\beta z}+{\cal{O}}(\text{csch}^2\beta z/\delta) .
\end{eqnarray}
Therefore, in the thin-wall limit 
\begin{equation}\label{FMThin}
\lim_{\delta\rightarrow 0}e^{-A(z)}=e^{\beta |z|}+\frac{\alpha}{\beta}\sinh(\beta z) ,
\end{equation}

Notice that for $z<0$ and $z>0$, (\ref{FMThin}) is a vacuum solution of the Einstein field equations, and
\begin{eqnarray}
    \lim_{\delta\rightarrow 0} &&G^b{}_a=6\beta\delta(z)\left(\partial t^a dt_b+\partial x_{i}^a dx^i_b\right)\nonumber\\&&\qquad+\left[6\alpha(2\beta-\alpha)\Theta(-z)-6\alpha(2\beta+\alpha)\Theta(z)\right]\nonumber\\&&\qquad \times\left(\partial t^a dt_b+\partial x_{i}^a dx^i_b+\partial z^a dz_b\right)\ .
\end{eqnarray}
This means that the spacetime $(\mathbb{R}^5, g)$, where $g$ is given by (\ref{Ap1}, \ref{Ap2}, \ref{FM+}), can be identified in the limit $\delta\rightarrow 0$ with the
spacetime $(\mathbb{R}^5, g)$, with $g$ given by (\ref{FMThin}), generated by a thin domain wall with energy-momentum tensor given by
\begin{eqnarray}
&& T^b{}_a=6\beta\delta(z)\left(\partial t^a dt_b+\partial x_{i}^a dx^i_b\right)\nonumber\\&&\qquad+\left[6\alpha(2\beta-\alpha)\Theta(-z)-6\alpha(2\beta+\alpha)\Theta(z)\right]\nonumber\\&&\qquad \times\left(\partial t^a dt_b+\partial x_{i}^a dx^i_b+\partial z^a dz_b\right)\ .
\end{eqnarray}

Finally, to be rigorous one should prove that the metric (\ref{Ap1}, \ref{Ap2}, \ref{FM+}) provides a sequence of metrics that satisfies the
convergence condition required in \cite{Geroch:1987qn} in order to relate the limit of the curvature tensor distributions
with the limit of the metric tensor, but it is not our concern here.

\end{document}